\documentclass[12pt,a4paper,twoside]{article}
\usepackage{epsfig}
\usepackage{baltlat6}
\usepackage{wrapfig}
\usepackage{graphicx}

\begin{document}
\titlehead{Baltic Astronomy ---- vol.YY, XXX--XXX, 2009.}
\titleb{Gamma-ray bursts: connecting the prompt emission with the afterglow}

\begin{authorl}
\authorb{P. Veres}{1,2} 
and
\authorb{Z. Bagoly}{1} 
\end{authorl}

\begin{addressl}
\addressb{1}{Dept. of Physics of Complex Systems, E\"otv\"os University, H-l117 Budapest, P\'azmány P. s. 1/A, email: veresp@elte.hu}
\addressb{2}{Dept. of Physics, Bolyai Military University, H-1581 Budapest, POB 15, Hungary}
\end{addressl}

\submitb{Received 2009. June}

\begin{abstract}
With the early afterglow localizations of gamma-ray burst positions made by
Swift, the clear delimitation of the prompt phase and the afterglow is not so
obvious any more. It is important to see weather the two phases have the same
origin or they stem from different parts of the progenitor system. We will
combine the two kinds of gamma-ray burst data from the Swift-XRT instrument
(windowed timing and photon counting modes) and from BAT. A thorough desription
of the applied procedure is given. We apply various binning techniques to the
different data: Bayes blocks, exponential binning and signal-to-noise type of
binning. We present a handful of flux curves and some possible applications.
\end{abstract}

\section{Introduction}
Gamma-ray bursts are the most energetic phenomenon in the Universe. After their
discovery (Klebesadel et al., 1973), and the detection of the first counterpart
in other wavelengths than gamma now the Swift satellite (Gehrels et al. 2004)
is almost routinely observing X-ray afterglows.  In the study of GRB prompt
emission and afterglow, it is a straightforward idea to combine flux curves of
GRBs from gamma-ray and X-ray data. A wide spectral coverage leads to a more
complete picture of the phenomenon. 
In our analysis have extrapolated the gamma-ray flux into the X-ray band to
have a commond ground for analysis. It can be done the other way around,
extrapolating the X-ray flux to the BAT energy range. Most bursts in our sample
come from the long and possibly the intermediate duration group (Horv\'ath
1998, Bal\'azs et al. 1998, Bal\'azs et al. 1999, Horv\'ath 2002, Horv\'ath et
al. 2008, \v{R}\'{\i}pa et al. 2009)

\section{Data reduction and binning}
Swift gathers $\gamma$-ray and X-ray data relevant for our analysis. We choose
three different approaches to bin the flux curves: Bayesian method for the
gamma-ray data, equal binning in logarithmic coordinates in the case of the
windowed timing (WT) XRT data and a signal-to-noise type of binning in the case
of photon counting (PC) XRT data. The flux curves and the spectra were
generated using standard \texttt{HEASoft} tools and the most recent calibration
database. Initial calibration was made using \texttt{xrtpipeline} and {\texttt
batgrbproduct} pipeline scripts with the latest calibration databases.

In gamma-rays the results of the primary pipeline processing contain, among
others, a $64$ ms resolution gamma-ray flux curve. This was used as the input
to the Bayesian block analysis. We cut our combined $15-150$ keV flux curve
using Bayes blocks as presented in Scargle (1998). We set a large prior for the
algorithm so it will stop at an early point (which corresponds to a small
number of change points) making sure that we have enough resolution for each of
the bins. After deducing the time intervals we have used {\texttt batbinevt} to
bin the data and get a raw spectrum (pha) file. The further steps recommended
in the BAT analysis thread were also carried out.  For each interval, we
created the appropriate response matrices and fitted both a power law and a
power law with a high-energy cutoff. We use the criterion from Sakamoto et al.
(2008) to choose between the two models. If the $\chi^2$ improves by more than
$6$ by using a cutoff power law, we use the latter instead of the simple power
law model. The next step is to extrapolate the model from the gamma-ray band
($15-150$ keV) into the X-ray band ($0.3$ - $10$ keV).

In X-rays the WT mode is active when the count rate of the source is high (over
$\sim 10$ counts/s). This means we have a good signal-to-noise ratio and we can
bin the counts in equal bins in logarithmic space. We have fitted a spectrum
for the whole duration of the WT mode and got a conversion factor from rates to
flux. There is a more detailed description of the procedure in the next
section.

For the PC mode we bin our data to have a signal-to-noise ratio of at least
$3$. We do this by incrementing the endpoint of our interval in time until the
count rate reaches the required level. At this point we store this interval and
repeat the procedure until the end of the observing period. We correct for the
pile-up in the detector as described in Vaughan et al. (2006) for the PC mode.
For the WT mode pileup correction is made according to Romano et al. (2008)\\

The next step is to convert the count rates to flux. To do this, we divide the
cumulative flux curve in $n$ parts, each with equal number of counts. $n$ is
chosen by hand depending on the intensity of the afterglow from $2$ to $6$.
For each time slice we fit a spectrum to get the count equivalent in
erg/cm$^2$. The spectra are fitted with \texttt{Xspec} using individual
anciliary response functions and the most recent response function available.
We used an absorbed power law model of the form: $$ (\mathrm{wabs}_G) \times
(\mathrm {zwabs}_S)\times \mathrm{(power law)}.$$ The absorbing column density
(NH) of the Milky Way was taken from Dickey et Lockman (1990) (denoted here by
$\mathrm{wabs}_G$), and, where redshift was known, the source absorbtion was
also fitted (denoted here by $\mathrm {zwabs}_S$). If the redshift was unknown
$\mathrm {zwabs}_S$ was substituted with a simple absorbing $\mathrm {wabs}_S$
component. The spectra were binned using \texttt{grppha} so all channels had
minimum $20$ counts.\\

Evans et al. (2007) use the same method, but they get the conversion factor
(erg/cm$^2$/s equivalent of $1$ count/s ) by integrating over the entire
spectrum (corresponding to $n=1$). We report here that our conversion factors
are in good agreement with those in Evans et al. (2007) and their related
web-page.

\section{Interesting cases}

Here we present several GRBs to illustrate our method of binning and combining
data. For now, we include only bursts from the long and intermediate duration
population. The behaviour of the afterglow requires the use of a log scale but
several GRBs have significant flux before the trigger (precursors with $t <
T_{trigger} = 0$). For this reason we choose to include a shift in the time
axis where necessary. 

\begin{figure} \includegraphics[width= {0.999\columnwidth} ]{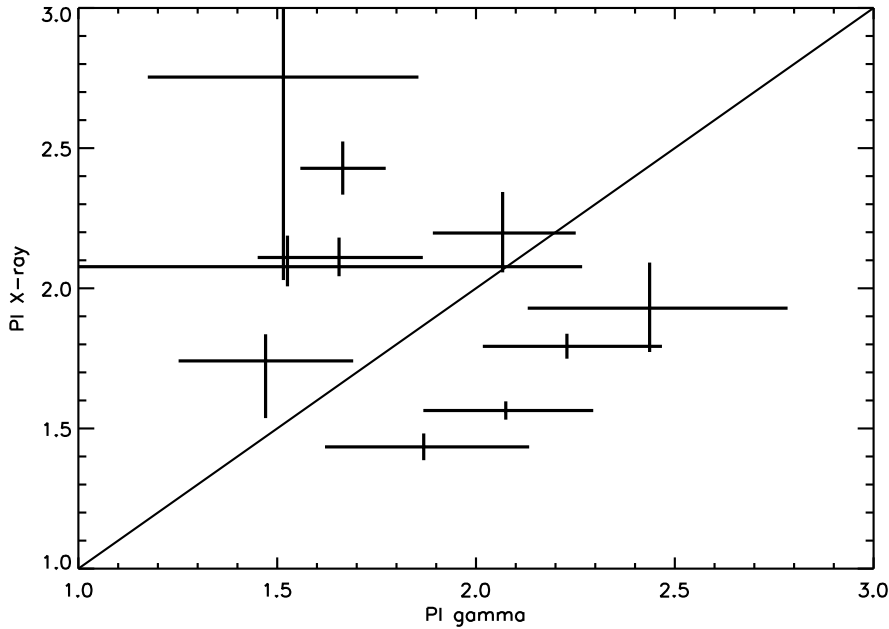}
\caption{Photon indices from the late prompt emission phase and the early
afterglow phase. The continuous line represents equality between the two
parameters} \label{fig:aa} \end{figure}

Another way of examining the problem is to consider gamma-ray and X-ray photon
indices. If the emission stems from the same population of electrons, we would
expect the two types of indices to be equal within errors. The measurements
should be carried out roughly at same time. We have taken a sample 10 bright
Swift bursts and compared the two indices. We found that there was a visible
discrepancy. (Figure \ref{fig:aa})

\vspace{0.5cm}

\begin{center}
\begin{figure}
  \includegraphics[width={0.99\columnwidth}]{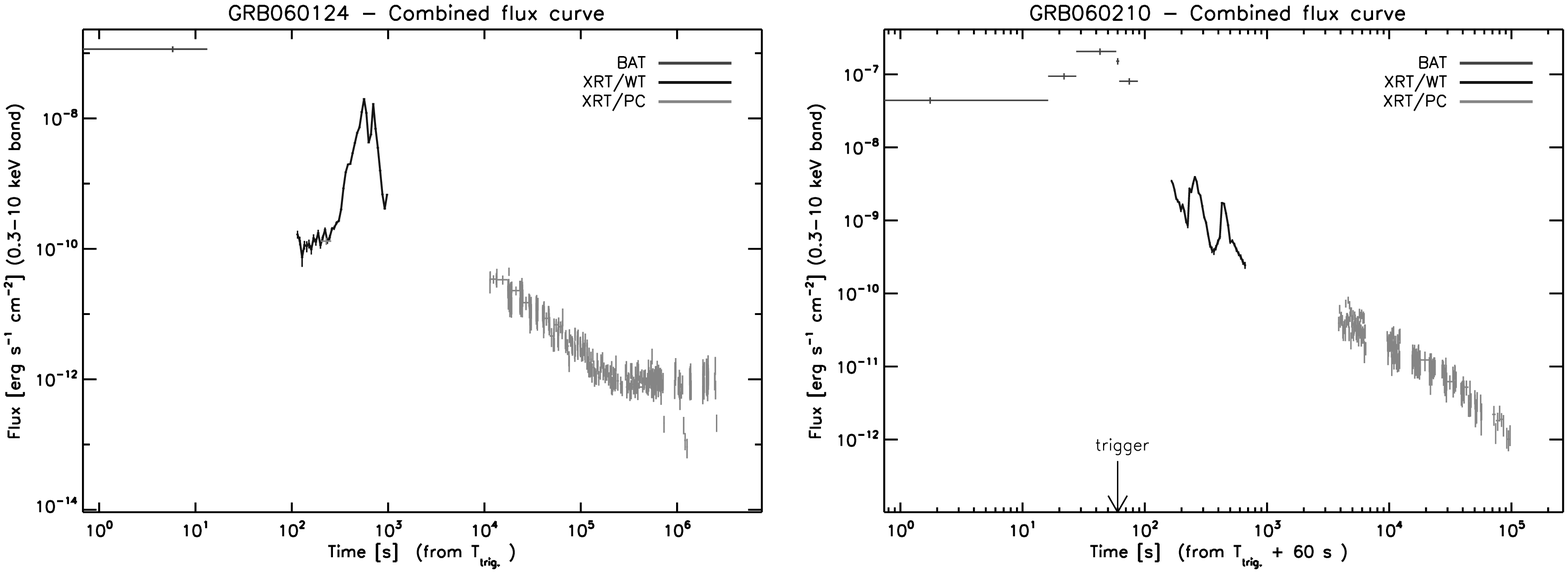}
\vspace{0.3cm}
\includegraphics[width= {0.99\columnwidth} ]{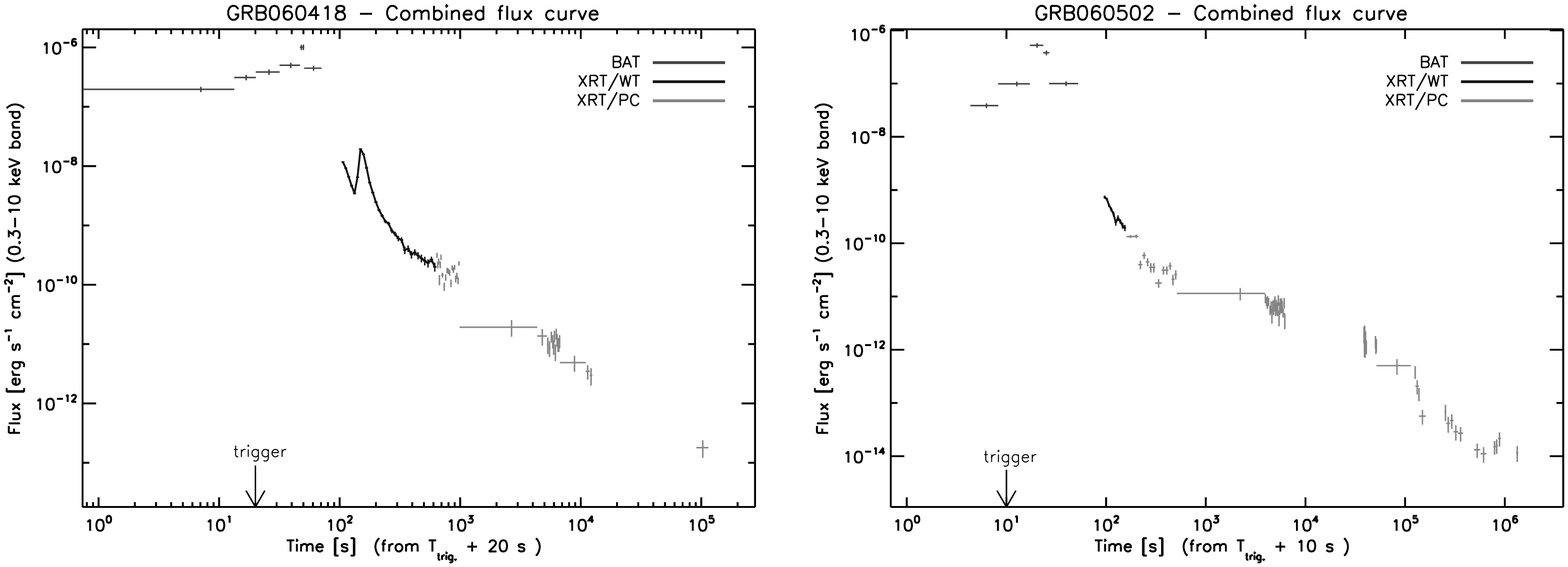}
\vspace{0.3cm}
  \includegraphics[width={0.99\columnwidth}]{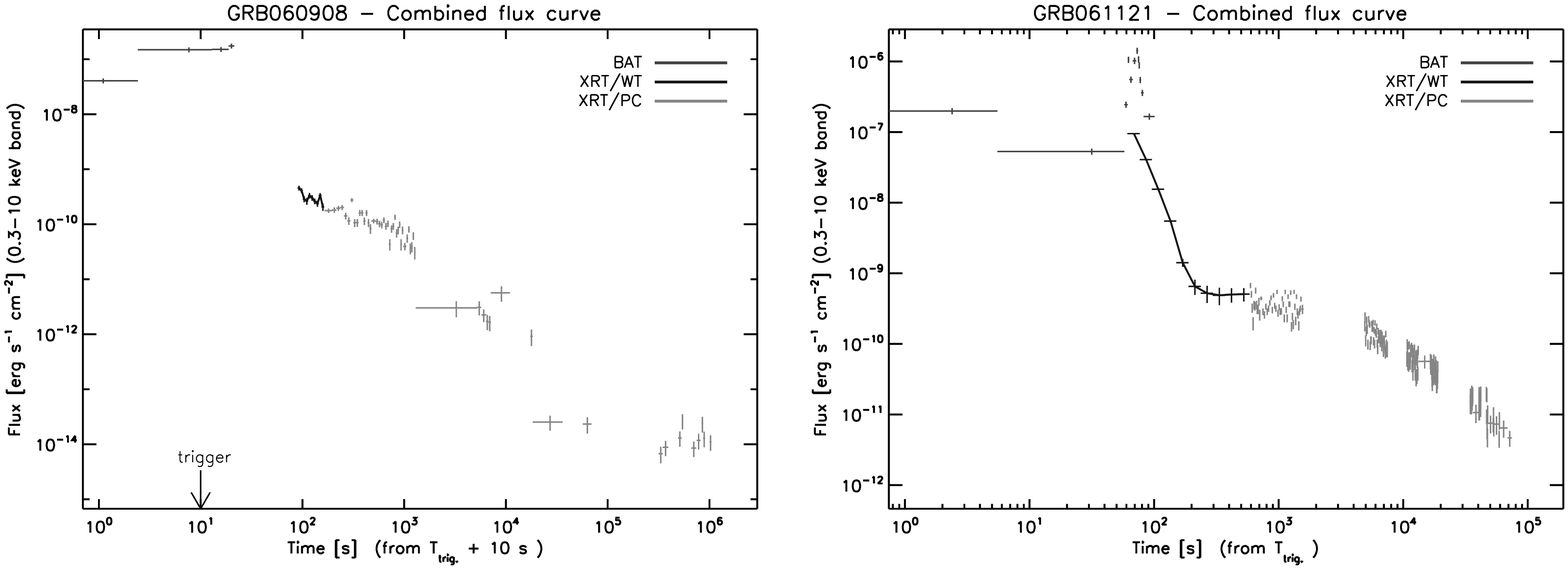}
\caption{A few examples of combined flux curves of Swift GRBs.}
\label{fig:ex}
\end{figure}
\end{center}

\section{Discussion and Conclusion} 

We have selected a few bright GRBs (Figure \ref{fig:ex}) to present our binning
method for data from two bands. There are cases where the possible
extrapolation of the WT mode data seems to match the gamma-ray flux curve (for
the first three GRBs: 060210, 060418 and 060502). This could be proof of a
connection between the processes which govern the prompt phase and the
processes of the early afterglow. It looks possible that the steep decline is
indeed a sequel of the prompt phase seen in X-rays. No such claim can be made
for the following two events (060604 and 060908) because of the scarceness of
the data. At the last two events (061121 and 060124) however, there is an
apparent discrepancy between the gamma-ray curve and the X-ray curve. This
discrepancy is best seen at GRB061121. This could mean the extrapolation of the
gamma-ray spectrum was not adequate.

ACKNOWLEDGMENTS. This research is supported from Hungarian OTKA grant K077795.

\References
\refb
Bal\'azs, L.G., M\'esz\'aros, A. and Horv\'ath, I.  1998, A\&A, 339, 1
\refb
Bal\'azs, L.G., M\'esz\'aros, A., Horv\'ath, I. and Vavrek, R., 1999, A\&AS, 138, 417
\refb 
Dickey, J.M. and Lockman, F.J., 1990, ARA\&A, 28, 215
\refb 
Evans, P.A., Beardmore, A.P., Page, K.L. et al., 2007, A\&A, 469, 379
\refb
Gehrels, N., Chincarini, G., Giommi, P. et al., 2004, ApJ 611, 1005 
\refb
Horv\'ath, I., 1998, ApJ, 508, 757 
\refb
Horv\'ath, I., 2002, A\&A, 392, 791 
\refb
Horv\'ath, I., 2009, Ap\&SS, 323, 83 
\refb 
Klebesadel, R.W., Strong, I.B. and Olson, R.A.,  ApJ, 182, 85
\refb
\v{R}\'{\i}pa, J.,  M\'esz\'aros, A., Huja, D.  2009, A\&A, 498, 399
\refb 
Romano, P., Campana, S., Chincarini, G. et al., 2006, A\&A, 456, 917 
\refb 
Sakamoto, T., Barthelmy, S.D., Barbier, L. et al., 2008, ApJ, 175, 179
\refb 
Scargle, J. D., 1998, ApJ, 504, 405
\refb 
Vaughan, S., Goad, M.R., Beardmore, A.P. et al., 2006, ApJ, 638, 920

\end{document}